\documentclass[aps]{revtex4}
\usepackage{graphicx}
\usepackage{amsmath}
\usepackage{amssymb}
\newcommand{\be}{\begin{equation}}
\newcommand{\ee}{\end{equation}}
\newcommand{\bea}{\begin{eqnarray}}
\newcommand{\eea}{\end{eqnarray}}

\newcommand{\bse}{\begin{subequations}}
\newcommand{\ese}{\end{subequations}}

\begin{document}
\title{A Critical Analysis of Goldberger-Wise Stabilization in Randall-Sundrum Scenario}
\author{ Anindya Dey\footnote{E-mail: anindya@mri.ernet.in}}
\affiliation{Department of Physics\\ 
Harish-Chandra Research Institute\\
Chhatnag Road, Jhusi, Allahabad - 211019, India}
\author{Debaprasad Maity\footnote{E-mail: tpdm@mahendra@iacs.res.in}, Soumitra SenGupta
\footnote{E-mail: tpssg@mahendra@iacs.res.in}}
\affiliation{Department of Theoretical Physics, Indian Association for the
Cultivation of Science,\\
Calcutta - 700 032, India}
\begin{abstract}
The Goldberger-Wise mechanism of stabilizing modulus in the Randall-Sundrum braneworld,by
introducing a bulk scalar field with quartic interaction terms localized at the
3-branes has been extremely popular as a stabilizing mechanism when the back-reaction 
of the scalar field on the geometry is negligibly small. In this note we re-examine the
mechanism by an exact analysis without resorting to the approximations adopted by Goldberger and
Wise. An exact calculation of the stabilization condition indicates the existence of closely spaced 
minimum and a maximum for the potential and also brings out some new features involved 
in the context of stabilization of such braneworld models.
\end{abstract}
\maketitle

The gauge hierarchy problem,which refers to the vast disparity between the
weak scale and the Planck scale, leads to the well-known naturalness problem 
in connection with the Higgs mass in the standard model of elementary particles. 
One can avoid such problem by invoking supersymmetry at the expense of 
including a large number of particles (namely superpartners) in the theory, all 
of which so far unseen.

Recently, braneworld models with large extra spatial dimensions have been 
suggested \cite{add} which show that a large extra dimensional volume may
bring down the fundamental mass scale of the theory to the TeV scale and thus resolves
the naturalness problem. In an alternative approach Randall and Sundrum (RS)
\cite{rs} proposed a higher dimensional scenario consisting of a small extra 
spatial dimension orbifolded as $S_1/Z_2$, with two opposite tension branes 
located at the two orbifold fixed points. With a negative five-dimensional 
bulk cosmological constant, they obtained the following space-time metric as a 
solution of the five dimensional Einstein's equation,
\begin{equation}
ds^2=e^{-2kr|\phi|}\eta_{\mu\nu}dx^{\mu}dx^{\nu} - r^2 d\phi^2
\label{ansatz}
\end{equation}
Such a warped geometry immediately resolves the naturalness problem by warping the
scalar masses from Planck scale in one brane (called Planck brane or hidden brane) 
to TeV scale in the other brane( called the visible brane or standard model brane) 
for very tiny value ($\sim$ near Planck length) of the modulus $r$ which measures 
the separation of the two branes. In this scenario, $r$ is associated with the 
vacuum expectation value (VEV) of a massless four-dimensional scalar field 
(modulus field) which has zero potential, so that $r$ is not determined by the 
dynamics of the model. Goldberger and Wise (GW) \cite{gw} proposed to generate such 
a potential classically by introducing a bulk massive scalar field with quartic 
interaction terms localized at the two 3-branes and finally obtained a value 
$kr \sim 12$ by minimizing the potential, without any fine-tuning of parameters. 
In this analysis the back-reaction of the scalar field on the background geometry was neglected.
A similar analysis by taking full back-reaction into consideration have been studied where the nature of the 
warped geometry as well as the possible stability of the modulus have been readdressed\cite{dmssgss}.
In the present note however we stick to the original GW approach where the back-reaction can indeed be ignored and re-examine 
the modulus stabilization mechanism more critically. We extend the analysis
beyond the approximations made in \cite{gw} and bring out the dependence of the stability condition
on the values of the various parameters. We show the existence of a maximum along with a minimum of the modulus
potential for appropriate values of the parameters leading to possible instability.\\

We begin with the Goldberger-Wise analysis, which specifies an 
action of the form:
\begin{subequations}
\begin{eqnarray}
S &=& S_{Gravity} + S_{vis} + S_{hid} + S_{scalar}~, \\
\mbox{where,} &&\nonumber \\
S_{Gravity} &=& \int d^4x~r~d{\phi} \sqrt{G}~[ 2M^3R + \Lambda]\\
S_{vis} &=&  \int d^4x \sqrt{-g_{s}}~[L_{s} - V_{s}]\\
S_{hid} &=&  \int d^4x \sqrt{-g_{p}}~[L_{p} - V_{p}]\\
S_{scalar}&=&\frac{1}{2}\int d^4x \int _{-\pi}^{\pi}\sqrt{G}(G^{AB}\partial_{A}
\Phi\partial_{B}\Phi - m^2 \Phi^2) -
\int \sqrt{-g_{p}}\lambda_{p}(\Phi^2-v_{p})^2- \int d^4x \sqrt{-g_{s}}
\lambda_{s}(\Phi^2 - v_{s})^2.
\end{eqnarray}
\end{subequations}
Here $\Lambda$ is the five dimensional cosmological constant, 
$V_{s}, V_{p}$ are the visible and hidden brane
tensions. \\
$\Phi$ develops a $\phi$-dependent vacuum expectation value, which has to be
determined from the classical equation of motion,
\begin{equation}
\partial_ \phi(e^{-4\sigma}\partial_\phi \Phi)=m^2 r^2 e^{-4\sigma}\Phi
+ 4 ^{-4\sigma}\lambda_{v} r \Phi(\Phi^2 - v_{s}^2)\delta (\phi-\pi)
+ 4 e^{-4\sigma}\lambda_{p} r \Phi(\Phi^2 - v_{p}^2)\delta (\phi)
\label{eqm}
\end{equation}\\
where $\sigma=kr|\phi|$.
In the bulk (where the delta functions are not relevant), the equation has a 
general solution:
\begin{equation} \label{Phi}
\Phi(\phi)=e^{2\sigma}[A e^{\nu\sigma} + B e^{-\nu\sigma}]
\end{equation}
where $\nu=\sqrt{4 + {m^2}/{k^2}}$.                                         
The solution is now plugged back into the original scalar field action 
and integrated over $\phi$ to obtain an effective 4-dimensional potential
for r, of the form:                         
\begin{equation}
V_{\Phi}(r)=k(\nu + 2)A^2 (e^{2\nu kr \pi} -1) + k(\nu - 2)B^2(1-e^{-2\nu kr \pi}) + 
\lambda_{s} e^{-4kr\pi}[\Phi(\pi)^2 -v_{s}^2]^2 + \lambda_{p}[\Phi(0)^2 - v_{p}^2].
\label{pot}
\end{equation}
The coefficients should be determined by matching the delta-function terms at 
the boundaries, as they appear in Eq.(\ref{eqm}). The resultant condition
on $A$ and $B$ is obtained as follows:
\begin{subequations}\label{bc}
\begin{eqnarray} 
k[(2+\nu)A + (2-\nu)B] -2\lambda_{p}\Phi(0)[\Phi(0)^2 - v_{p}^2]=0.\label{bd1}{\label{bd1}}\\
k e^{2kr\pi}[(2+\nu)e^{\nu kr\pi}A + (2-\nu)e^{-\nu kr \pi}B] + 
2\lambda_{s}\Phi(\pi)[\Phi(\pi)^2 - v_{s}^2] = 0.
\label{bd2}
\end{eqnarray}
\end{subequations}
At this stage The authors of \cite{gw} have argued that in the large $\lambda$
limit, one can choose $\Phi(0)=v_{p}$ and $\Phi(\pi)=v_{s}$ as the minimum 
energy configuration. However clearly with this choice both the Eq.(\ref{bc}) 
can only be satisfied  for a non-vanishing value of $r$ only if $\lambda$ is infinitely
large.

Before re-examining the analysis of \cite{gw} for arbitrary value of $\lambda$ ,we first calculate the first and second 
derivative of the potential to find the exact extremization condition for the modulus. 
It may be noted that such conditions were obtained in \cite{gw} using some specific approximations in the
large $\lambda$ limit. Our result here is exact and perfectly general for all values of $\lambda$. 
A long but straightforward calculation yields, the first derivative for the potential (assuming
explicit $r$ dependence of $A$ and $B$ through the Eq.(\ref{bc}) as,
\begin{equation} {\label{V'}}
V'_{\Phi}(r)= -4k \pi \lambda_{s} e^{-4kr\pi}(\Phi(\pi)^2 - v_{s}^2)^2
-4k^2\pi[(\nu+2)e^{2\nu kr\pi}A^2 + (2-\nu)e^{-2\nu kr\pi}B^2 + (4-\nu^2)AB].
\end{equation}
where prime denotes differentiation with respect to $r$.

Using the extremization condition for the 
potential $(V'_{\Phi}(r) = 0)$ from the above equation we obtain  
a simple but exact form for the second derivative of the potential as
\begin{equation}{\label{V''}}
V''_{\Phi}(r)= 4 k^2 \pi \nu [{(2+\nu)AB' + (\nu - 2)BA'}]
\end{equation}
The sign of R.H.S. of the above equation determines exactly whether the stationary value of the
modulus $r$ is a stable value or not.

Assume that for arbitrary value of $\lambda$ ( not infinity) the boundary value of the scalar field at the two orbifold fixed
points are $\Phi(\phi = 0) = Q_p(r)$ and $\Phi(\phi = \pi) = Q_s(r)$. Now
we can easily express the undetermined constants $A$ and $B$ in terms of the quantities,
$Q_p$ and  $Q_s$ as,
\begin{eqnarray} \label{AB}
A = \frac {Q_s(r) e^{-2 \sigma} - Q_p(r) e^{- \nu \sigma}}{ 2 \sinh(\nu \sigma)} \\
B = \frac {Q_p(r) e^{\nu \sigma} - Q_s(r) e^{- 2 \sigma}}{ 2 \sinh(\nu \sigma)}
\end{eqnarray}
Substituting the expressions for $A$ and $B$ in Eq.\ref{bc} we obtain,
\begin{eqnarray} 
\frac {\nu} { \sinh(\nu \sigma)} \left [e^{- 2 \sigma} - \{\frac{2 + \nu}{2\nu} e ^{- \nu \sigma}
+ \frac{\nu -2}{2\nu} e^{ \nu \sigma}\}\frac {Q_p}{Q_s}\right] = \frac {2 \lambda_p}{k}
\frac {Q_p}{Q_s}(Q_p^2 - v_p^2) 
\label{bcQp}\\
 \frac {\nu} {\sinh(\nu \sigma)} \left [\frac {Q_p}{Q_s} - \{\frac{2 + \nu}{2\nu}
  e ^{(\nu -2) \sigma}
+ \frac{\nu -2}{2\nu} e^{-(\nu+2) \sigma}\}\right] = \frac {2 \lambda_p}{k}
 (Q_s^2 - v_s^2)e^{- 2 \sigma}
 \label{bcQs} 
 \end{eqnarray}

Further using Eq.\ref{bcQs} and the expressions for $A$ and $B$ into equ.\ref{V'} under stability condition we arrive at ( for $\lambda_s \neq 0$),
\begin{equation} \label{root}
x^2\left( 1 + \frac k {\lambda_s Q_s^2}\right) = \tilde{C}^2,
\end{equation}
where 
\begin{equation}
x = \frac {Q_p}{Q_s} - \frac{2 + \nu}{2 \nu} e ^{(\nu -2) \sigma}
- \frac {\nu -2}{2 \nu} e^{-(\nu+2) \sigma},~~~~~~~\tilde{C} = \left\{\frac{2 + \nu}{2 \nu} e ^{(\nu -2) \sigma} +\frac {\nu -2}{2 \nu} e^{-(\nu+2) \sigma}\right\} C
\end{equation}

Now, it is easy to manipulate the expression given below from the Eq.\ref{root} that is
\begin{equation}{\label{kr}}
 k r = \frac 1 {\pi (\nu -2)} \ln \left[\left(\frac 1 { \frac {2 + \nu}{2 \nu} + 
 \frac {\nu - 2}{2 \nu}
 e^{-2 \nu \sigma}}\right) \left(\frac {Q_p(r)}{Q_s(r)}\right) 
 \left(\frac 1 {1 \pm C \sqrt \frac {\lambda_s 
 Q_s(r)^2}{k + \lambda_s Q_s(r)^2}}\right)
 \right]  
\end{equation}
for the stationarity condition i.e $V'_{\Phi}(r) = 0 $. 
Here
\begin{equation}
 C = \sqrt{ 1 - \frac{ \frac 4 {\nu} \{(2+\nu)e^{2 (\nu -2) \sigma} - 
 e^{- 4 \sigma} (4-\nu^2) + (2 - \nu)e^{-2 (\nu + 2) \sigma}\}}{\{(\nu + 2)e^{(\nu -2)\sigma}
 + (\nu - 2)e^{-(\nu +2)\sigma}\}^2}} 
\end{equation}
It may be noted that in the large $k r$ limit, $C \sim \sqrt{\frac {\nu -2 }{\nu +2 }}$. 
The expression for $k r $ in equ.\ref{kr}), is an exact expression for the stationary value of the
modulus $r$ and is valid for any value of the brane coupling constants $\lambda$.

Using the above results, the expression for $V''$ becomes
\begin{equation} 
V_{\Phi}''(r) = - \frac {4 k \pi \nu e^{- 2 \sigma}}  {\sinh{(\nu \sigma)}} 
\left[ \lambda_p(Q_p^2 - v_p^2)Q_p Q_p'
+ \lambda_s(Q_s^2 - v_s^2)Q_s Q_s' + \pi \lambda_p\lambda_s(Q_s^2 - v_s^2)(Q_p^2 - v_p^2)Q_p Q_s
\right]
\label{V''2}
\end{equation}
where 'prime'$\{'\}$ denotes derivative with respect to $r$.

At this point, we want to emphasize that so far no approximation has been made and
all the results are exact.
\vskip .2cm
{ \it \underline{Stability Analysis}}\\

We now re-examine the stability of the modulus $r$.

{\bf Case I:}~~~~~~ $\lambda \rightarrow \infty $

From the expression for the potential Eq.\ref{pot}, the minimum energy configuration
leads to the 
\begin{equation}
Q_s \rightarrow v_s ~~~;~~~ Q_p \rightarrow v_p
\end{equation}
which are constants. This completely agrees withe result of \cite{gw}. However some
interesting new features also show up in this analysis which we discuss now.

In this limit the expressions for the stationary points become(in the large $kr$ limit)
\begin{equation}
k r = \frac 1 {\pi (\nu -2)} \ln\left[\frac {2 \nu}{2+\nu} \frac {v_p(r)}{v_s(r)} 
\frac 1 {1 \pm \sqrt{\frac {\nu -2 }{\nu +2 }}}
\right] 
\end{equation}

In the infinity limit of the coupling constant
\begin{equation}
Q_s' = 0, ~~~~~~~;~~~~~~~~~ Q_p' = 0
\end{equation}

So, the Eq.\ref{V''2} becomes
\begin{equation}
V_{\Phi}''(r) = - \frac {4 k \pi \nu e^{- 2 \sigma}}  {\sinh{(\nu \sigma)}} 
\left[ \pi \lambda_p\lambda_s(Q_s^2 - v_s^2)(Q_p^2 - v_p^2)Q_p Q_s
\right]
\end{equation}

Now, equ.\ref{bcQp} tells that in a wide range of parameter values, the expression
 $\lambda_p (Q_p^2 - v_p^2)$ is negative where as 
 from the Eq.\ref{bcQs}, $\lambda_p (Q_p^2 - v_p^2)$ is either 
positive or negative according to the value of $k r$. However, for  
\begin{equation}
k r_{+} = \frac 1 {\pi (\nu -2)} \ln\left[\frac {2 \nu}{2+\nu} \frac {v_p(r)}{v_s(r)} 
\frac 1 {1 + \sqrt{\frac {\nu -2 }{\nu +2 }}}
\right],
\end{equation}
$V''(r) > 0$, that means $k r_{+}$ is a stable point for the minimum of the potential.
So, for the other value of $k r_{-}$, the potential will be maximum.
Clearly no extreme fine tuning of the parameters is required to get the right magnitude 
for $k r$ . 
Here, one important point is that the expression for the value of $ k r$ is different
from that of \cite{gw}. This is why if we calculate
the value of $k r $ corresponding to the values of parameters $m/k = 0.2$ and 
$v_p/v_s = 1.5$ given in \cite{gw} we get $k r = 10.846$ which is lesser than
what authors have predicted. 
But, for instance, taking $v_p/v_s = 2.3$ and $\nu = 2.02$ yields $k r = 12.2504$ for 
the minimum of the potential.
The value of $kr = 14.4993$ corresponds to maximum of the potential. 
Clearly these two values are very close to each other.\\ 

{\bf Case II:}~~~~~~ $\lambda$  is finite but very large, 

From the Eq.\ref{bc}, it is clear that if the brane coupling constant is large but finite, the value of the
scalar field $Q_p$ on the Planck brane is lower than $v_p$ and approaches $v_p$ as the value of the brane coupling 
constant tends to infinity.
But for the standard model brane it is just the opposite where the scalar field value on the brane 
$Q_s$, approaches $v_s$ from above as the brane coupling constant approaches infinity from a finite value.
This prompts us to calculate the corrections to the boundary scalar field values
in the leading $1/{\lambda}$ order correction. These become
\begin{eqnarray} \label{modQ}
Q_p(r) = v_p + \frac {k v_s}{\lambda_p v_p^2}  \frac {\nu e^{- 2  \sigma}}{4 \sinh{(\nu\sigma)}}
\left[\frac {v_s}{v_p} - \left\{\frac {2 + \nu}{2\nu} e^{(2 -\nu) \sigma} + \frac {\nu - 2}
{2 \nu} e^{(\nu + 2) \sigma}\right\}\right] \nonumber\\
Q_s(r) = v_s + \frac {k v_p} {\lambda_s v_s^2}  \frac {\nu e^{2 \sigma}}{4 \sinh{(\nu\sigma)}}
\left[\frac {v_p}{v_s} - \left\{\frac {2 + \nu}{2\nu} e^{(\nu-2) \sigma} + \frac {\nu - 2}
{2 \nu} e^{-(\nu + 2) \sigma}\right\}\right]
\end{eqnarray}
\begin{figure}
\hfill
\begin{minipage}{.48\textwidth}
\includegraphics[width=3.20in,height=3.0in]{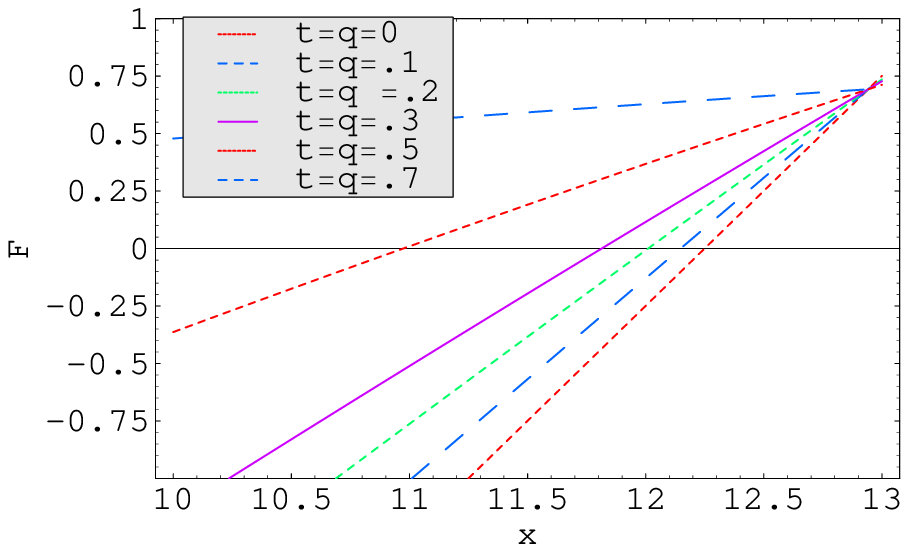}
\caption{F$(kr,\nu, n, t, q)$ with $kr$ taking the different set of values
for the parameters $t$ and $q$ but keeping fixed values of $\nu = 2.02 $ and $n = 1.5$} \label{fig1}
\end{minipage}
\hfill
\begin{minipage}{.48\textwidth}
\includegraphics[width=3.20in,height=3.0in]{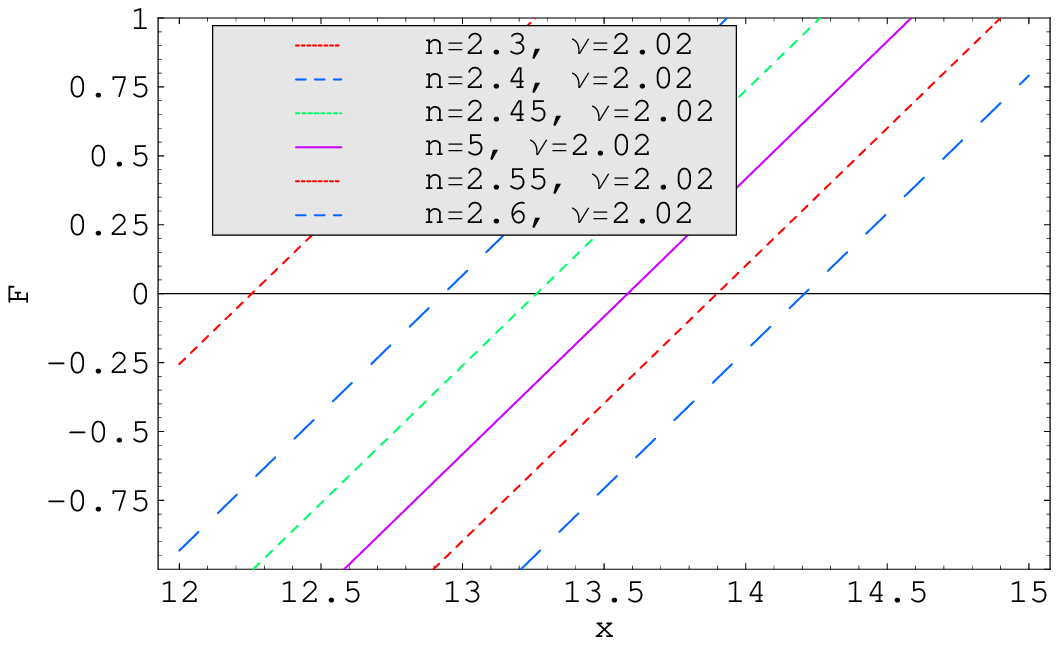} 
\caption{Plot of F$(kr,\nu, n, t, q)$ with $kr$ taking the different set of values
for the parameters $n$ and $\nu$ but keeping fixed values of $t = 0.001 $ and $q = 0.001$} \label{fig2}~~~~~~~~~~~~~
\end{minipage}
\hfill
\end{figure}
Now, we are interested in studying the modified expression for Eq.\ref{kr} under  
large but finite values of brane coupling constants upto first order in $1/\lambda$ correction.
A straightforward calculation using Eqs.\ref{modQ},leads to the value, (in the large $k r$ limit)
\begin{equation} \label{krMod}
kr = \frac 1 {\pi (\nu -2)} \ln\left[\frac {2 \nu}{2+\nu} 
\frac n {1 \pm \sqrt{\frac {\nu -2 }{\nu +2 }}(1 - \frac 1 2 q)} \left\{1 + \frac {\nu -2 }
{2\nu} e^{(2 - \nu) k r \pi} \frac {\nu} 8 \frac t n - \frac {\nu} 8 q n \left(n - \frac {\nu + 2 }
{2\nu} e^{(\nu- 2 ) k r \pi}\right)\right\}\right] 
\end{equation}
where, $\nu, n = v_p/v_s, t = k/(\lambda_p v_p^2)$ and $ q = k/(\lambda_s v_s^2)$ are four
parameters of the model.Here also as in the previous case we obtain one minimum as well as one maximum for the potential. 
The positive sign corresponds to the  value of $kr$ which minimizes
the radion potential.
For this positive sign,
the Eq. \ref{krMod} reduces to
F$(kr,\nu,n,t,q) = 0$, where F is an appropriate function of the parameters. 

We plot the function F with
respect to $kr$ for different values of the parameters $n,\nu,t,q$ as shown in the figures \ref{fig1},\ref{fig2} and find
out the roots of F$(kr,\nu,n,t,q)$ in the region where the value of $kr$ solves the hierarchy problem i.e. $kr \sim 12$. 
The existence of nontrivial roots in this region is clearly shown in the above figures.\\
Our analysis therefore brings out the following new features in the context of modulus stabilization:\\
While Godberger-Wise calculation calculation was strictly correct only for infinite value of $\lambda$     
our analysis here shows that this mechanism stabilizes the modulus $r$ even for finite but large value of $\lambda$.
The stable value of the modulus in this case once again solves the hierarchy problem without any unnatural fine tuning of the parameters
although the stable value for the modulus differs marginally from that estimated by GW.
We have determined the modified value of $kr$ ,inflicted from the finiteness of $\lambda$ ,to the leading order correction 
around the infinite value i.e. in terms of inverse of $\lambda$.  
As stated earlier our exact calculation indicates that there exists simultaneously
a very closely spaced ($\sim$ Planck length$(l_p)$) maximum along with the minimum.This may have interesting consequences 
in a quantum mechanical version of such a model which however is beyond the scope of interest of this note.

\vspace{.1cm}
\noindent
{\bf Acknowledgment}\\
DM acknowledges Council of Scientific and Industrial Research, Govt. of 
India for providing financial support. Authors acknowledge stimulating discussions with 
J.K.Bhattacharjee, K.Ray and S.Sen.

\end{document}